\documentclass{aa}

\usepackage{rotating}
\usepackage{natbib}
\usepackage{mathtools}
\usepackage{amsmath}

\usepackage[percent]{overpic}

\DeclarePairedDelimiter\abs{\lvert}{\rvert}%
\makeatletter
\let\oldabs\abs
\def\abs{\@ifstar{\oldabs}{\oldabs*}}

\begin{document}

\title{Intermediate shock substructures within a slow-mode shock occurring in partially ionised plasma}
\titlerunning{Intermediate shock substructure}

\author{B. Snow\inst{1} and A. Hillier\inst{1}}
\institute{University of Exeter, Exeter, EX4 4QF, UK \email{b.snow@exeter.ac.uk} \label{inst1}}


\abstract 
{Slow-mode shocks are important in understanding fast magnetic reconnection, jet formation and heating in the solar atmosphere, and other astrophysical systems. The atmospheric conditions in the solar chromosphere allow both ionised and neutral particles to exist and interact. Under such conditions, fine substructures exist within slow-mode shocks due to the decoupling and recoupling of the plasma and neutral species. 
}
{We study numerically the fine substructure within slow-mode shocks in a partially ionised plasma, in particular, analysing the formation of an intermediate transition within the slow-mode shock.}
{High-resolution 1D numerical simulations are performed using the (P\underline{I}P) code using a two-fluid approach.}
{We discover that long-lived intermediate (Alfv\'en) shocks can form within the slow-mode shock, where there is a shock transition from above to below the Alfv\'en speed and a reversal of the magnetic field across the shock front. 
The collisional coupling provides frictional heating to the neutral fluid, resulting in a Sedov-Taylor-like expansion with overshoots in the neutral velocity and neutral density. The increase in density results in a decrease of the Alfv\'en speed and with this the plasma inflow is accelerated to above the Alfv\'en speed within the finite width of the shock leading to the intermediate transition. This process occurs for a wide range of physical parameters and an intermediate shock is present for all investigated values of plasma-$\beta$, neutral fraction, and magnetic angle. As time advances the magnitude of the magnetic field reversal decreases since the neutral pressure cannot balance the Lorentz force. The intermediate shock is long-lived enough to be considered a physical structure, independent of the initial conditions.}
{Intermediate shocks are a physical feature that can exist as shock substructure for long periods of time in partially ionised plasma due to collisional coupling between species.}


\maketitle

\section{Introduction}

Shocks occur readily in the lower solar atmosphere, driven by wave steepening, e.g., umbral flashes \citep{Beckers1969,Houston2018}, or magnetic reconnection driven events, such as Ellerman bombs \citep{Ellerman1917,Nelson2013}. The listed phenomena occur in regions of the sun where partially ionised effects are thought to play a key role in the underlying physics. As such, to understand phenomena occurring in the lower solar atmosphere, we must also understand the role of partial ionisation in shocks. 

In magnetohydrodynamics (MHD), there are three characteristic wave speeds (slow, Alfv\'en and fast) which leads to a multitude of potential shock transitions \citep[see for example][]{Delmont2011}. The slow-mode shock (transition from super-slow to sub-slow flow speeds) is of particular importance due to its role in magnetic reconnection, where a change in connectivity of the magnetic field results in a release of stored magnetic energy. The energy can be released instantaneously via Joule heating, or post-reconnection by the influence of the magnetic field on the plasma. The classical 2D schematic for fast reconnection features slow-mode shocks \citep{Petschek1964}. Recent work suggests more complicated reconnection configurations occur in the solar atmosphere, e.g., plasmoid instability, 3D topology, extended MHD \citep[see reviews by][]{Yamada2010,Pontin2011,Loureiro2016,Cassak2017}. In the Petschek model, the presence of slow-mode shocks in the outflow region allow efficient transport of energy away from the reconnection region, increasing the reconnection rate. Slow-mode shocks are also found to form in alternative models for fast reconnection \citep{Liu2012,Innocenti2015,Shibayama2015}.

The intermediate shock (transition from above to below the Alfv\'en velocity) has been shown to exist in single-fluid resistive MHD representative of the Earth's magnetopause \citep{Karimabadi1995}. The physical mechanism being that the non-ideal region around the shock allows a separation of the magnetic field and fluid, and the formation of an intermediate shock between the two ideal regions. The stability of intermediate shocks was proved by \cite{Wu1991} who studied numerically the formation of these shocks due to wave steepening. The relationship between intermediate shocks and slow-mode shocks was shown analytically by \cite{Hau1989} for resistive MHD shocks, finding a relation between the downstream slow-mode speed and the strength of the intermediate transition.

In partially ionised plasma, such as in the solar chromosphere or prominances, shocks become more complex. \cite{Hillier2016} studied the formation of slow-mode shocks in partially ionised plasmas. Around the shock front, the plasma and neutral species decouple and recouple resulting in a finite width slow-mode shock. 
The Lorentz force indirectly affects neutrals through collisions and hence the drift between ion and neutral species becomes an important parameter in partially-ionised plasma. However, observing the ion-neutral drift has been found to be difficult due to line-of-sight effects \citep{Anan2017}. A review of partially-ionised modelling can be found in \cite{Khomenko2017}. 


In this paper, we utilise high-resolution two-fluid numerical simulations to investigate the substructure within slow-mode shocks. We discover that long-lived intermediate shocks can form within the finite width of the shock as a result of the fluid coupling and decoupling around the shock front. This result has applications to the solar chromosphere, as well as interplanetary and interstellar partially ionised plasma. We study the formation of compound intermediate shocks in partially ionised plasma, and the conditions in which such shock structures can form. 


The outline of this paper is as follows. First we define the shock transitions and classifications to be used in this paper, and the analytical solution to the magnetohydrodynamic (MHD) and partially ionised plasma (PIP) equations in the Hoffman-Teller shock frame. Next, the numerical methods and initial conditions under consideration are introduced. Following this, a reference case is presented where we compare an MHD and a PIP simulation, and identify the key differences, investigating the physical evolution that produces the intermediate shock. Finally, we consider a parameter study and investigate the effect of the plasma properties on the lifetime and magnitude of the shock. We find that for our initial conditions, an intermediate shock will always form in a partially-ionised plasma. However the lifetime and magnitude depend heavily on the plasma properties.

\section{Methodology}

\subsection{Shock classifications}



MHD waves have three characteristic speeds: Alfv\'en ($V_A$), slow ($V_s$), and fast ($V_f$). As such, multiple shock transitions are possible, depending on the magnitude of the velocity, relative to the characteristic speeds in the pre- and post-shock regions.

Following the approach of \cite{Delmont2011}, we classify shock transitions using the relationship between the normal flow velocity $v_{\perp}$ and the characteristic speeds: 
\begin{itemize}
\item (1) superfast: $V_f < \abs{v_\perp}$,
\item (2) subfast: $V_A < \abs{v_\perp} < V_f $,
\item (3) superslow: $V_s < \abs{v_\perp} < V_A$,
\item (4) subslow: $0< \abs{v_\perp} < V_s$,
\item ($\infty$) static: $v_\perp = 0$.
\end{itemize}
Defining the upstream condition $\text{u}$ and downstream condition $\text{d}$, several shocks of the form $\text{u} \rightarrow \text{d}$ are possible:
\begin{itemize}
\item $1 \rightarrow 2 $ fast shocks,
\item $3 \rightarrow 4$ slow shocks,
\item $1 \rightarrow 2=3$ switch-on,
\item $2=3 \rightarrow 4$ switch-off,
\item $1\rightarrow 3, 1 \rightarrow 4, 2 \rightarrow 3, 2 \rightarrow 4$ intermediate shocks.
\end{itemize}


\subsection{Hoffman-Teller Equations}

The Hoffman-Teller frame allows for jump relations to be derived from the MHD equations. In this choice of rest frame, the velocity and magnetic field vectors are in the same plane either side of the shock, i.e., the electric field across the shock is zero. This reduces the MHD equations to a 2-dimensional problem for variables perpendicular ($\perp$) and parallel ($\parallel$) to the shock front.

\subsubsection{MHD Solution}

\begin{figure}
    \centering
    \includegraphics[width=0.95\linewidth,clip=true,trim=7.4cm 5.0cm 7.5cm 7.8cm]{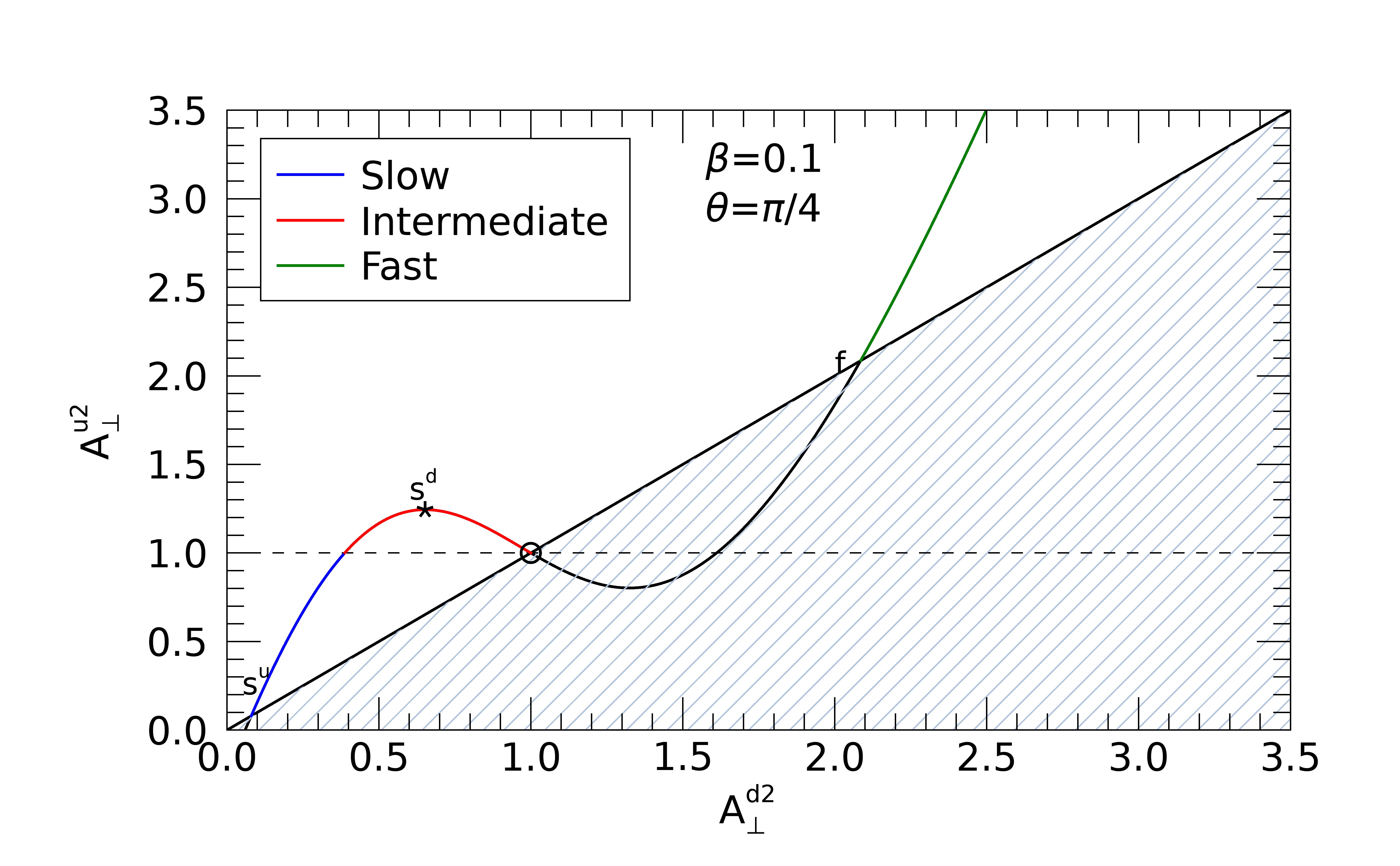}
    \caption{Hau-Sonnerup shock solution for plasma-$\beta ^{\text{u}} =0.1$ and $\theta ^{\text{u}} =0.4$. Shaded region shows impossible transitions. Possible transitions include: slow-mode shock (blue), intermediate shock (red), rotational discontinuity (circle), and fast-mode shocks (green).}
    \label{fig:HT-mhd}
\end{figure}


In the Hoffmann-Teller frame, the MHD equations can be integrated and the upstream ($^{\text{u}}$) and downstream ($^{\text{d}}$) conditions can be equated:
\begin{gather}
\rho^{\text{u}} v_\perp ^{\text{u}} = \rho^{\text{d}} v_\perp ^{\text{d}}, \label{eqn:htmhd1}\\
\rho ^{\text{u}} v_\perp ^{\text{u}} v_\parallel ^{\text{u}} - \frac{1}{\mu _0} B_\perp ^{\text{u}} B_\parallel ^{\text{u}} = \rho ^{\text{d}} v_\perp ^{\text{d}} v_\parallel ^{\text{d}} - \frac{1}{\mu _0} B_\perp ^{\text{d}} B_\parallel ^{\text{d}}, \\
\rho^{\text{u}} v_\perp^{\text{u}} v_\perp^{\text{u}} + P^{\text{u}} +\frac{B^{{\text{u}}2}}{2 \mu_0} = \rho^{\text{d}} v_\perp^{\text{d}} v_\perp^{\text{d}} + P^{\text{d}} +\frac{B^{{\text{d}}2}}{2 \mu_0}, \\
\frac{\gamma}{\gamma -1} \frac{P^{\text{u}}}{\rho^{\text{u}}} + \frac{v^{\text{u}2}}{2} = \frac{\gamma}{\gamma -1} \frac{P^{\text{d}}}{\rho^{\text{d}}} + \frac{v^{\text{d}2}}{2}, \\
v_\perp^{\text{u}} B_\parallel^{\text{u}} -v_\parallel^{\text{u}} B_\perp^{\text{u}} = v_\perp^{\text{d}} B_\parallel^{\text{d}} -v_\parallel^{\text{d}} B_\perp^{\text{d}} \\
B_\perp^{\text{u}} = B_\perp^{\text{d}}. \label{eqn:htmhd2}
\end{gather}
We can also define the following: Alfv\'en Mach number as $A_\perp ^2 = v_\perp ^2 \frac{\mu_0 \rho}{B_x^2}$, plasma-$\beta$ as $\beta = \frac{2 \mu_0 P}{B^2}$, and the angle between the velocity and magnetic field $\theta$.

A solution to Equations (\ref{eqn:htmhd1}-\ref{eqn:htmhd2}) can be found in \cite{Hau1989} relating the upstream Alfv\'en velocity to the upstream plasma-$\beta$, upstream $\theta$, and downstream Alfv\'en velocity, i.e.,

\begin{gather}
    A_\perp ^{\text{u}2} = \left[ A_\perp ^{\text{d}2} \left( \frac{\gamma-1}{\gamma} \left( \frac{\gamma+1}{\gamma -1} -\tan ^2 \theta ^{\text{u}} \right) \left(A_\perp ^{\text{d}2} -1 \right) ^2 \right. \right. \nonumber\\ 
    + \left. \left. \tan ^2 \theta ^{\text{u}} \left( \frac{\gamma-1}{\gamma} A_\perp ^{\text{d}2} -1 \right) \left(A_\perp ^{\text{d}2} -2 \right) \right) - \frac{\beta ^{\text{u}}}{ \cos ^2 \theta ^{\text{u}}} \left( A_\perp ^{\text{d}2} -1 \right) ^2 \right]  \nonumber\\
    / \left[ \frac{\gamma -1}{\gamma} \frac{\left( A_\perp ^{\text{d}2}-1 \right) ^2}{ \cos ^2 \theta ^{\text{u}}} - A_ \perp ^{\text{d}2} \tan ^2 \theta ^{\text{u}} \left( \frac{\gamma -1}{\gamma} A_\perp ^{\text{d}2} -1 \right) \right]. \label{eqn:hau}
\end{gather}

This is shown graphically in Figure \ref{fig:HT-mhd} for a choice of plasma-$\beta ^{\text{u}} =0.1$ and $\theta ^{\text{u}} =\pi /4$. The trivial solution is that the upstream and downstream velocities are identical, i.e., no shock transition. The shaded region shows impossible solutions, where the velocity is higher downstream than upstream. The non-trivial solutions show the possible shock transitions for given plasma-$\beta$ and $\theta$ values. This curve intersects the $A^{\text{u}}=A^{\text{d}}$ line at three points. The intersect labelled $s^{\text{u}}$ denotes the upstream slow speed. At the point $A_\perp ^{\text{u}} = A_\perp ^{\text{d}} =1$, a rotational discontinuity is possible, where there is a change in the angle of the magnetic field but plasma properties remain the same. Slow-mode transitions (blue line) are bounded by  $A_\perp ^{\text{u}} =1$.  The downstream slow speed (point labelled on Figure \ref{fig:HT-mhd} as $s^{\text{d}}$) corresponds to the critical value that separates strong ($2 \rightarrow 4$) and weak ($2 \rightarrow 3$) intermediate transitions. 
The strong transition ($2 \rightarrow 4$) is closely linked to the slow-mode transition ($3 \rightarrow 4$) through its relation to the downstream slow speed. 

\subsubsection{PIP Solution}

In the PIP equations, collisional terms are taken into account and therefore, the equations have a non-zero right-hand side, see Equations (\ref{eqn:neutral1}-\ref{eqn:plasma2}). As such, when the equations are integrated in the Hoffmann-Teller frame, there are integral terms on the right hand side, see Equations (\ref{app1:pip1}-\ref{app1:pip2}), as opposed to constants in the MHD Equations. 
However, by adding the neutral and plasma species together, the integral terms can be eliminated, and by choosing a point upstream and downstream such that the drift velocity equals zero, one recovers the MHD equations in the Hoffman-Teller frame, where the Alfv\'en speed and plasma-$\beta$ values depend on the \textbf{total} plasma density ($\rho_{\text{n}}+\rho_{\text{p}}$), see Appendix \ref{app:pipht}. These equations are also independent from the neutral fraction and hence, in the steady state solution, the values either side of the shock are governed by the MHD solution, when points are chosen such that the drift velocity is zero. However, within the shock, the species decouple and recouple hence in partially-ionised plasmas, it is of interest to study the substructure that occurs within the finite-width of the shock. 

\subsection{Numerical methods}

Two-fluid numerical simulations are performed using the (P\underline{I}P) code \citep{Hillier2016} which solves the interactions of a neutral fluid, and a coupled ion electron plasma. The simulations are 1D and use high resolution to resolve the substructure present within the shock. A first order HLLD (Harten-Lax-van Leer-Discontiunities) scheme is used to prevent spurious oscillations from occurring around the shock interface. The normalised PIP equations are given below:
\begin{gather}
\frac{\partial \rho _{\text{n}}}{\partial t} + \nabla \cdot (\rho _{\text{n}} \textbf{v}_{\text{n}})=0, \label{eqn:neutral1} \\
\frac{\partial}{\partial t}(\rho _{\text{n}} \textbf{v}_{\text{n}}) + \nabla \cdot (\rho _{\text{n}} \textbf{v}_{\text{n}} \textbf{v}_{\text{n}} + P_{\text{n}} \textbf{I}) \nonumber \\  \hspace{0.5cm} = -\alpha _c \rho_{\text{n}} \rho_{\text{p}} (\textbf{v}_{\text{n}}-\textbf{v}_{\text{p}}), \\
\frac{\partial e_{\text{n}}}{\partial t} + \nabla \cdot \left[\textbf{v}_{\text{n}} (e_{\text{n}} +P_{\text{n}}) \right] \nonumber \\  \hspace{0.5cm}= -\alpha _c \rho _{\text{n}} \rho _{\text{p}} \left[ \frac{1}{2} (\textbf{v}_{\text{n}} ^2 - \textbf{v}_{\text{p}} ^2)+ 3 \left(\frac{P_n}{\rho_n}-\frac{1}{2}\frac{P_p}{\rho_p}\right) \right], \\
e_{\text{n}} = \frac{P_{\text{n}}}{\gamma -1} + \frac{1}{2} \rho _{\text{n}} v_{\text{n}} ^2, \label{eqn:neutral2}\\
\frac{\partial \rho _{\text{p}}}{\partial t} + \nabla \cdot (\rho_{\text{p}} \textbf{v}_{\text{p}}) = 0 \label{eqn:plasma1}\\
\frac{\partial}{\partial t} (\rho_{\text{p}} \textbf{v}_{\text{p}})+ \nabla \cdot \left( \rho_{\text{p}} \textbf{v}_{\text{p}} \textbf{v}_{\text{p}} + P_{\text{p}} \textbf{I} - \textbf{B B} + \frac{\textbf{B}^2}{2} \textbf{I} \right) \nonumber \\  \hspace{0.5cm}= \alpha _c \rho_{\text{n}} \rho_{\text{p}}(\textbf{v}_{\text{n}} - \textbf{v}_{\text{p}}), \\
\frac{\partial}{\partial t} \left( e_{\text{p}} + \frac{\textbf{B}^2}{2} \right) + \nabla \cdot \left[ \textbf{v}_{\text{p}} ( e_{\text{p}} + P_{\text{p}}) -  (\textbf{v}_p \times \textbf{B}) \times \textbf{B} \right] \nonumber \\  \hspace{0.5cm} = \alpha _c \rho _{\text{n}} \rho _{\text{p}} \left[ \frac{1}{2} (\textbf{v}_{\text{n}} ^2 - \textbf{v}_{\text{p}} ^2)+ 3 \left(\frac{P_n}{\rho_n}-\frac{1}{2}\frac{P_p}{\rho_p}\right) \right],\\
\frac{\partial \textbf{B}}{\partial t} - \nabla \times (\textbf{v}_{\text{p}} \times \textbf{B}) = 0, \\
e_{\text{p}} = \frac{P_{\text{p}}}{\gamma -1} + \frac{1}{2} \rho _{\text{p}} v_{\text{p}} ^2, \\
\nabla \cdot \textbf{B} = 0,\label{eqn:plasma2}
\end{gather}
for neutral (subscript $\text{n}$) and plasma (subscript $\text{p}$) species. The neutral equations (\ref{eqn:neutral1}-\ref{eqn:neutral2}) are independent of the magnetic field. The plasma equations (\ref{eqn:plasma1}-\ref{eqn:plasma2}) are similar to the MHD equations however include collisional terms that couple the plasma to the neutral fluid. The collisional coefficient $\alpha _c$ is defined as
\begin{equation}
    \alpha _c = \alpha _0 \sqrt{\frac{T_n + T_p}{2 T_0}}
\end{equation}
where $T_0$ is the normalisation temperature such that the sound speed $c_{s}^2 = V_A ^2$ and, in the normalised form, $\alpha _0 = 1$. 
The magnetic field is normalised using $\textbf{B}=\textbf{B}/\sqrt{\mu_0}$. Details of the equations and their implementation in the (P\underline{I}P) code can be found in \cite{Hillier2016}. 

Our simulation data is translated into the shock frame by calculating the shock propagation speed in the MHD solution ($v_s$). The position and velocity variables are remapped to the shock frame, i.e., $v_{x {\text{n}}s,x{\text{p}}s}=v_{x{\text{n}},x{\text{p}}}-v_s$, $x_s = x/t - v_s$.

\begin{figure*}
\begin{overpic}[width=0.95\linewidth,clip=true,trim=0.9cm 7.8cm 1.5cm 7.8cm]{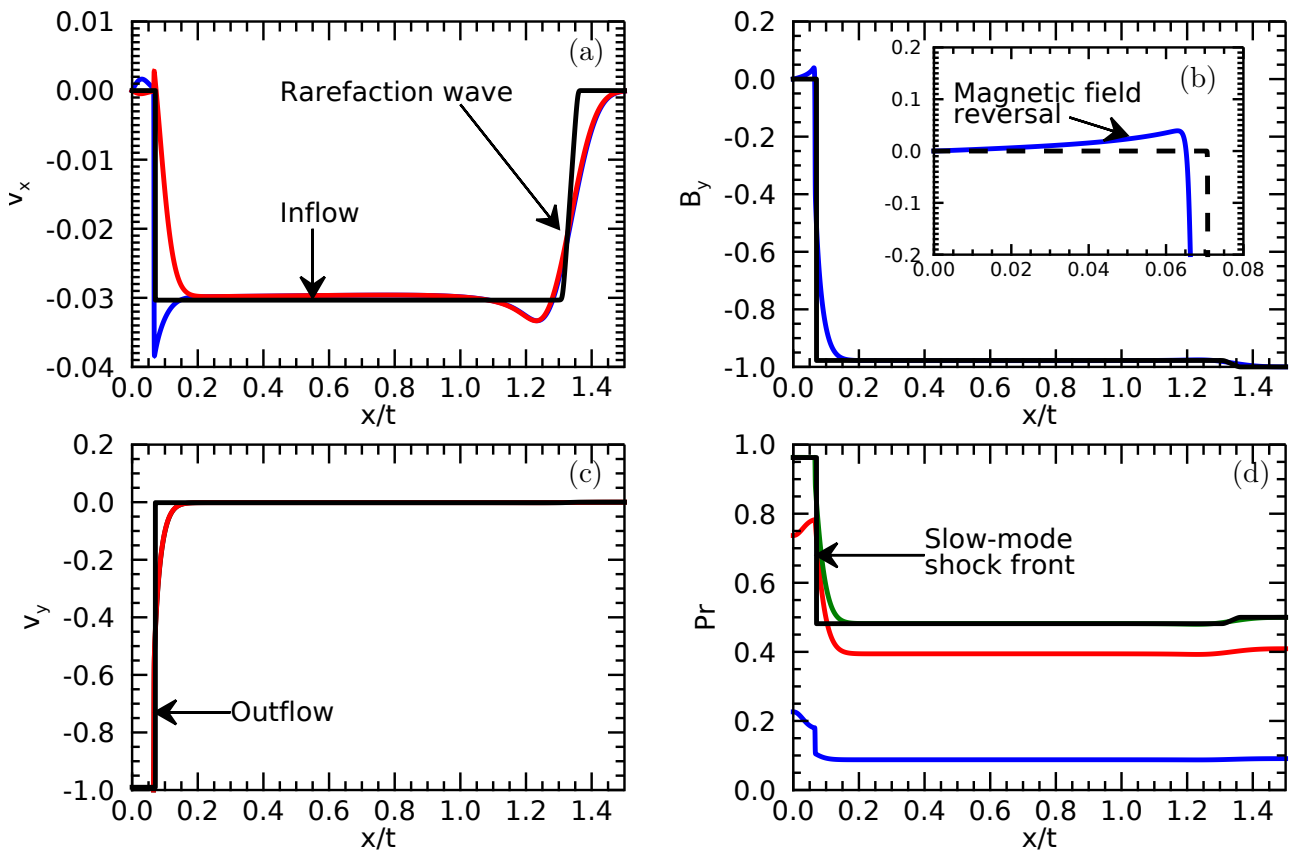}
 \put (42,60) {\large(a)}
 \put (88,58) {\large(b)}
 \put (42,28) {\large(c)}
 \put (92,28) {\large(d)}
\end{overpic}
\caption{MHD (black) and PIP (red neutral, blue plasma) reference solution for $\beta = 1.0$, $B_x=0.1$, $\xi _{\text{n}} =0.9$. (a) $v_x$ velocity. (b) $B_y$ magnetic field. (c) $v_y$ velocity. (d) Pressure. Green line indicates total pressure ($P_{\text{n}} +P_{\text{p}}$). The PIP solution is plotted after 2500 collisional times.}
\label{fig:bothref}
\end{figure*}

\subsection{Initial conditions}

The initial conditions in this paper are an extension of the model for slow-mode shock created in reconnection proposed by \cite{Petschek1964}. Initially, a discontinuous magnetic field is specified across the boundary. The plasma and neutral fluids are assumed to be in thermal equilibrium at the start of the simulation. 

\begin{gather}
B_x = 0.1 \\
B_y = -1.0 (x>0), 1.0 (x<0) \\
\rho _{\text{n}} = \xi _{\text{n}} \rho _{\text{t}} \\
\rho _{\text{p}} = \xi _i \rho _{\text{t}} \textbf{=} (1- \xi _{\text{n}}) \rho _{\text{t}} \\
P_{\text{n}} = \frac{\xi _{\text{n}}}{\xi_{\text{n}} + 2 \xi _i} P_{\text{t}} =  \frac{\xi _{\text{n}}}{\xi_{\text{n}} + 2 \xi _i} \beta \frac{B_0 ^2}{2} \\
P_{\text{p}} = \frac{2 \xi _i}{\xi_{\text{n}} + 2 \xi _i} P_{\text{t}} =  \frac{2 \xi _i}{\xi_{\text{n}} + 2 \xi _i} \beta \frac{B_0 ^2}{2}
\end{gather}
where $\xi _n$ and $\xi_i$ are the neutral and ion fractions respectively.

The $x=0$ boundary is treated as reflective, but such that magnetic field can penetrate the boundary. The system is normalised to have an Alfv\'en velocity of unity. The collisional time as determined by the bulk fluid density is calculated as $\tau = 1/(\alpha (T_0) \rho _{\text{t}}) = 1$. 

These initial and boundary conditions were used in the work of \cite{Hillier2016} to investigate the slow-mode shock. Here we investigate the formation and lifetimes of a feature that was present in Figures 4a and 5a of \cite{Hillier2016} but not discussed, namely the intermediate shock transition, which features a reversal of magnetic field across the shock interface.

\section{Results}

In this section, we present an MHD simulation and a PIP simulation and highlight the key differences. A snapshot of the results is shown in Figure \ref{fig:bothref}. Both simulations use 256,000 grid cells to resolve the spatial dimension. These two simulations use the following parameters: $\beta = 1.0$, $B_x = 0.1$, $\xi _{\text{n}} =0.9$. Note that the neutral fraction is only used in the PIP case and the MHD simulation is fully ionised. The effect of different parameters is investigated in Section \ref{sec:par}.  




\subsection{Reference MHD solution}



In the MHD case (Figure \ref{fig:bothref}, black lines), the initial conditions produce a rarefaction wave that drives fluid at the local Alfv\'en speed towards a slow-mode shock. The solution is expanding in time, however, is steady-state when plotted on a time-normalised axis ($x/t$), hence can be described as pseudo-steady.   


It is important to note that in the MHD solution, the slow-mode shock possesses no complex substructure, has a discontinuous jump (no finite width), and is purely a $3 \rightarrow 4$ transition. There is no reversal in magnetic field across the shock (see Figure \ref{fig:bothref}). These features do not hold when the model is extended to include additional effects (e.g. partial ionisation) where substructure can form in the shock front affecting the shock dynamics and resultant heating and energy transport. 

\subsection{Reference PIP solution}

The red and blue lines in Figure \ref{fig:bothref} show the neutral and ion fluid respectively for the PIP simulation, using the same parameters as the MHD case. The snapshot is taken after 2500 collisional times when the system can be assumed to be in a quasi-self-similar state. In the PIP case, the ion and neutral species decouple and recouple around the shock front, resulting in a finite width slow-mode shock where shock sub-structure can occur (compared to the discontinuous jump in the MHD case).

The initial conditions are a discontinuity in the magnetic field only, therefore directly affect the plasma only. The neutral fluid can only be influenced indirectly by collisions with the ionised plasma. As such, the system tends towards a pseudo-steady-state as time tends towards infinity. \cite{Hillier2016} presented the dynamic evolution for a similar case:

\begin{itemize}
    \item \textbf{Initialisation} The initial conditions are a discontinuity in the magnetic field and hence directly affect the plasma only, and the neutral fluid is affected via collisions. At $\tau =1 $ the system is highly decoupled. A Sedov-Taylor-like expansion occurs in the neutral fluid due to frictional heating between the two species.
    \item \textbf{Weak coupling} A fast-mode rarefaction wave forms and drives fluid towards a slow-mode shock. At this time, the rarefaction wave and slow-mode shock front are coupled and interact. 
    \item \textbf{Strong coupling} In this phase, the rarefaction wave and slow-mode shock decouple. The slow-mode shock can be considered to be independently evolving.
    \item \textbf{Quasi-self-similar state} System tends towards a self-similar state as time tend towards infinity. This is plotted in Figure \ref{fig:bothref}. The rarefaction wave is separated from the slow-mode shock, Figure \ref{fig:bothref}a. The two fluids are reasonably well coupled except within the slow-mode shock where large drift velocities are present, Figure \ref{fig:bothref}a. The neutral overshoot is present at late times indicating that it is a stable feature of the system (Figure \ref{fig:bothref}a).
\end{itemize}

A feature present but not discussed in the work of \cite{Hillier2016} is the reversal of the magnetic field across the shock front (Figure \ref{fig:bothref}b). This is a long-lived (but transient) feature of the system that exists long after the rarefaction wave and slow-mode shock have separated. The magnetic field reversal is a signature that an intermediate transition exists within the slow-mode shock.

\section{Analysis}


\subsection{Intermediate Shock}

\begin{figure*}
\begin{overpic}[width=0.95\linewidth,clip=true,trim=0.8cm 5.2cm 1.5cm 5.2cm]{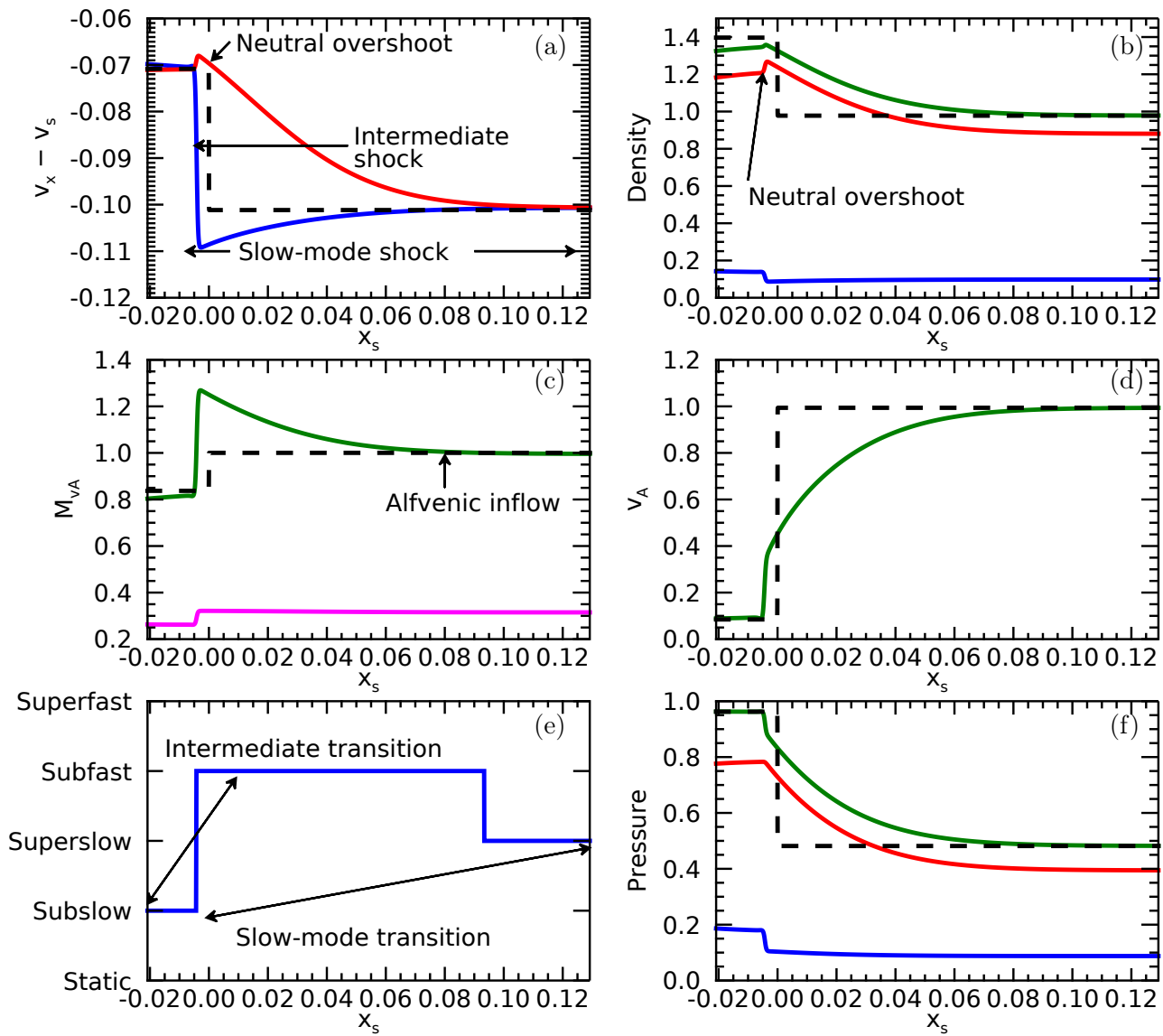}
 \put (44,85) {\large(a)}
 \put (92,85) {\large(b)}
 \put (44,57) {\large(c)}
 \put (92,57) {\large(d)}
 \put (44,28) {\large(e)}
 \put (92,28) {\large(f)}
\end{overpic}
\caption{MHD (black dashed) and PIP (red neutral, blue plasma) solutions within the shock. The figures are in the shock frame ($x_s = x/t -v_{s}$, where $v_{s}$ is the propagation speed of the shock in the MHD solution). (a) perpendicular velocity in the shock frame $v_x -v_{s}$. (b) Density. Green shows the total ($\rho_{\text{n}} + \rho_{\text{p}}$) density. (c) Alfv\'en Mach number using the plasma density (magenta) and total density (green). (d) Alfv\'en speed using the total density. (e) Shock transitions. (f) Pressure. Green line shows total pressure ($P_{\text{n}} + P_{\text{p}}$).}
\label{fig:isform}
\end{figure*}

\begin{figure}
\includegraphics[width=0.95\linewidth,clip=true,trim=0.8cm 7.6cm 1.0cm 8.0cm]{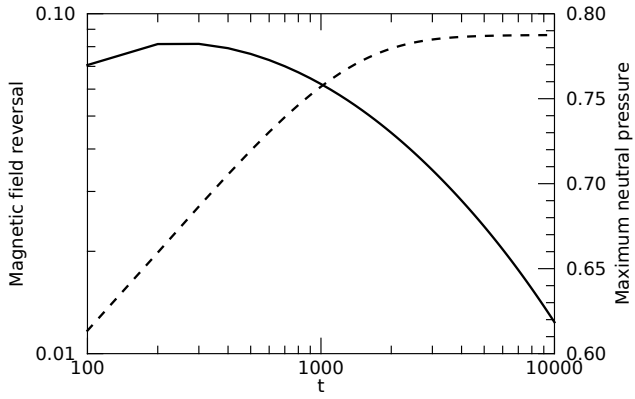}
\caption{Maximum neutral pressure (dashed) and maximum magnetic field reversal (solid) through time. Both quantities are maximum near the intermediate transition.}
\label{fig:decay}
\end{figure}

\begin{figure}
\includegraphics[width=0.95\linewidth,clip=true,trim=0.9cm 7.8cm 0.9cm 8.2cm]{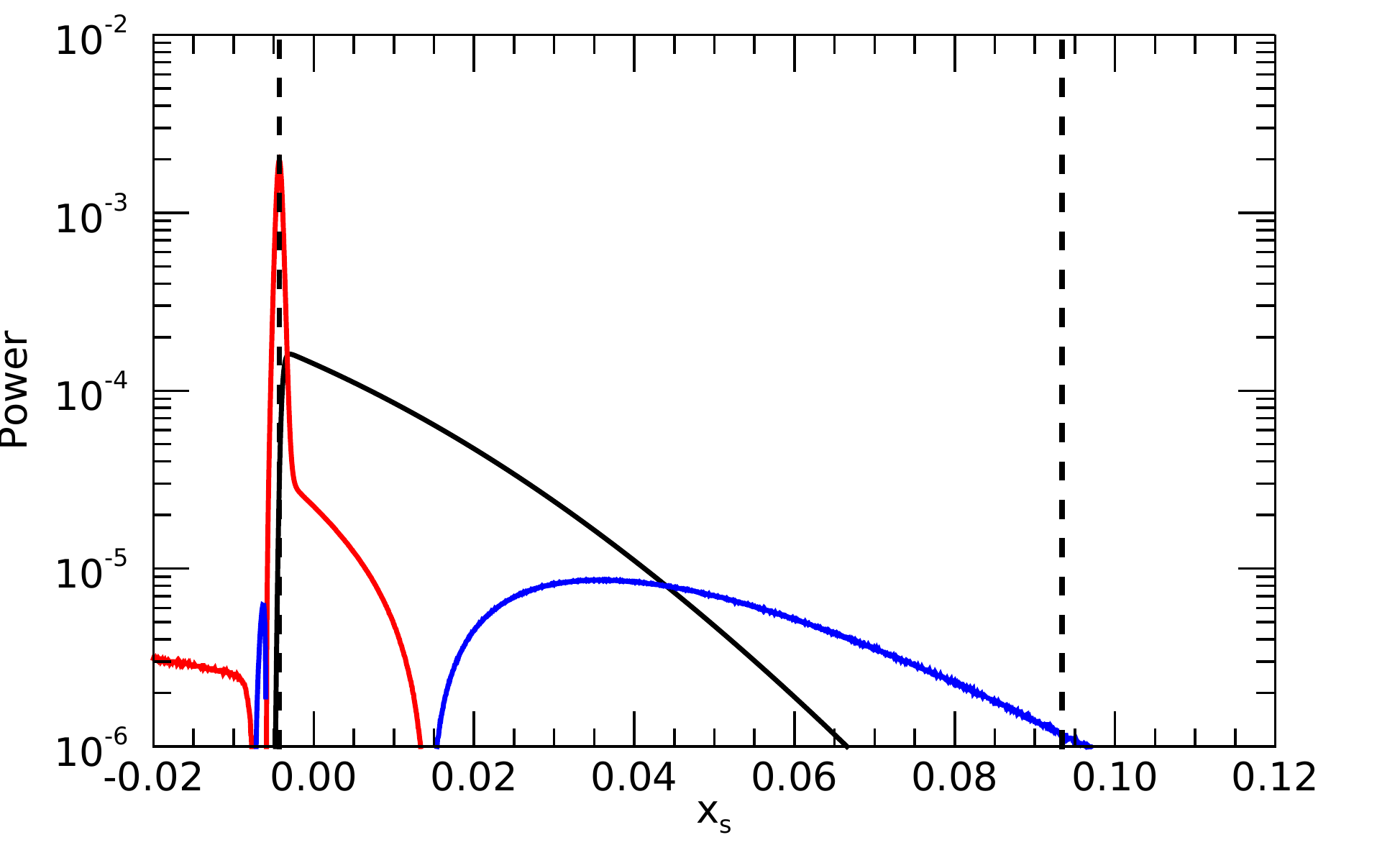}
\caption{Frictional heating (black line) and thermal damping (red and blue lines). The red line indicates ions losing heat to neutrals, and vice versa for the blue line. Dashed lines indicate the finite width of the shock.}
\label{fig:therm}
\end{figure}

The reversal in magnetic field in Figure \ref{fig:bothref}b is a key indicator of an intermediate shock, where the plasma transitions from above the local Alfv\'en speed, to below it. An intermediate shock is present in all partially ionised simulations as a transient feature of the system, however is not present in the MHD cases. The feature arises due to the interactions of ion and neutral particles. To analyse this feature, a new frame of reference is chosen such that the shock is stationary, i.e., $x_s = x/t - v_{s}$ where $v_{s}$ is the propagation speed of the MHD shock. In this frame, the velocity is also adjusted to account for the shock propagation speed, i.e., $v_x - v_{s}$. The shock frame allows us to correctly compute the transitions across the shock front and compare with the analytical results. The evolution of this structure is as follows:

\paragraph{(1) Sedov-Taylor-like expansion of the neutral fluid:} The initial conditions are a discontinuity in the magnetic field that affects the plasma only, resulting in a separation of neutral and plasma species in the shock front (Figure \ref{fig:isform}a). The large drift velocity creates a rapid heating of the neutral fluid. Subsequently, a hydrodynamic, Sedov-Taylor-like expansion occurs in the neutral fluid, resulting in an overshoot of neutral velocity (Figure \ref{fig:isform}a), and a sudden increase in neutral density (Figure \ref{fig:isform}b). 

\paragraph{(2) Acceleration of plasma:} The rarefaction wave drives velocity towards the slow-mode shock at the Alfv\'en speed (Figure \ref{fig:isform}c). This is also true in the MHD case and is a feature of the shock problem being studied. Within the slow-mode shock front, the bulk Alfv\'en speed decreases due to the increase in neutral density (Figure \ref{fig:isform}d) and the reversal in the magnetic field (Figure \ref{fig:bothref}b). The plasma inside the shock is accelerated to above the Alfv\'en speed (Figure \ref{fig:isform}c), resulting in an intermediate transition (Figure \ref{fig:isform}e).

\paragraph{(3) Decay with time:} The intermediate shock is sustained by the neutral pressure (Figure \ref{fig:isform}f), which, because of the finite coupling, cannot balance the Lorentz force. As time tends to infinity, the neutral pressure equalises and the magnetic field reversal becomes decreases in magnitude, see Figure \ref{fig:decay}. This is an indicator that the intermediate shock disappears with time. For these parameters, the magnetic field reversal is still prominent after 10000 collisional times, indicating that whilst this is a transient feature of the system, it is sufficiently long-lived to be physical and independent of the initial conditions. 

The frictional heating ($\alpha _c (T_{\text{n}},T_{\text{p}}) \rho_{\text{n}} \rho_{\text{p}} (v_{\text{n}}-v_{\text{p}})^2$) and thermal damping ($3 \alpha _c (T_{\text{n}},T_{\text{p}}) \rho_{\text{n}} \rho_{\text{p}} (P_{\text{n}}/\rho_{\text{n}}-P_{\text{p}}/(2\rho_{\text{p}}))$) between the two species are shown in Figure \ref{fig:therm} through the finite-width of the shock. Both the frictional heating and thermal damping reach their maximum values at the intermediate transition. Here, there is the largest drift velocity (Figure \ref{fig:isform}a). Either side of the shock, the species are reasonably well coupled and hence there is minimal frictional heating. 




\subsection{Parameter Study} \label{sec:par}

\subsubsection{Changing ionisation fraction}

\begin{figure}
\includegraphics[width=0.95\linewidth,clip=true,trim=0.9cm 8.0cm 0.9cm 8.0cm]{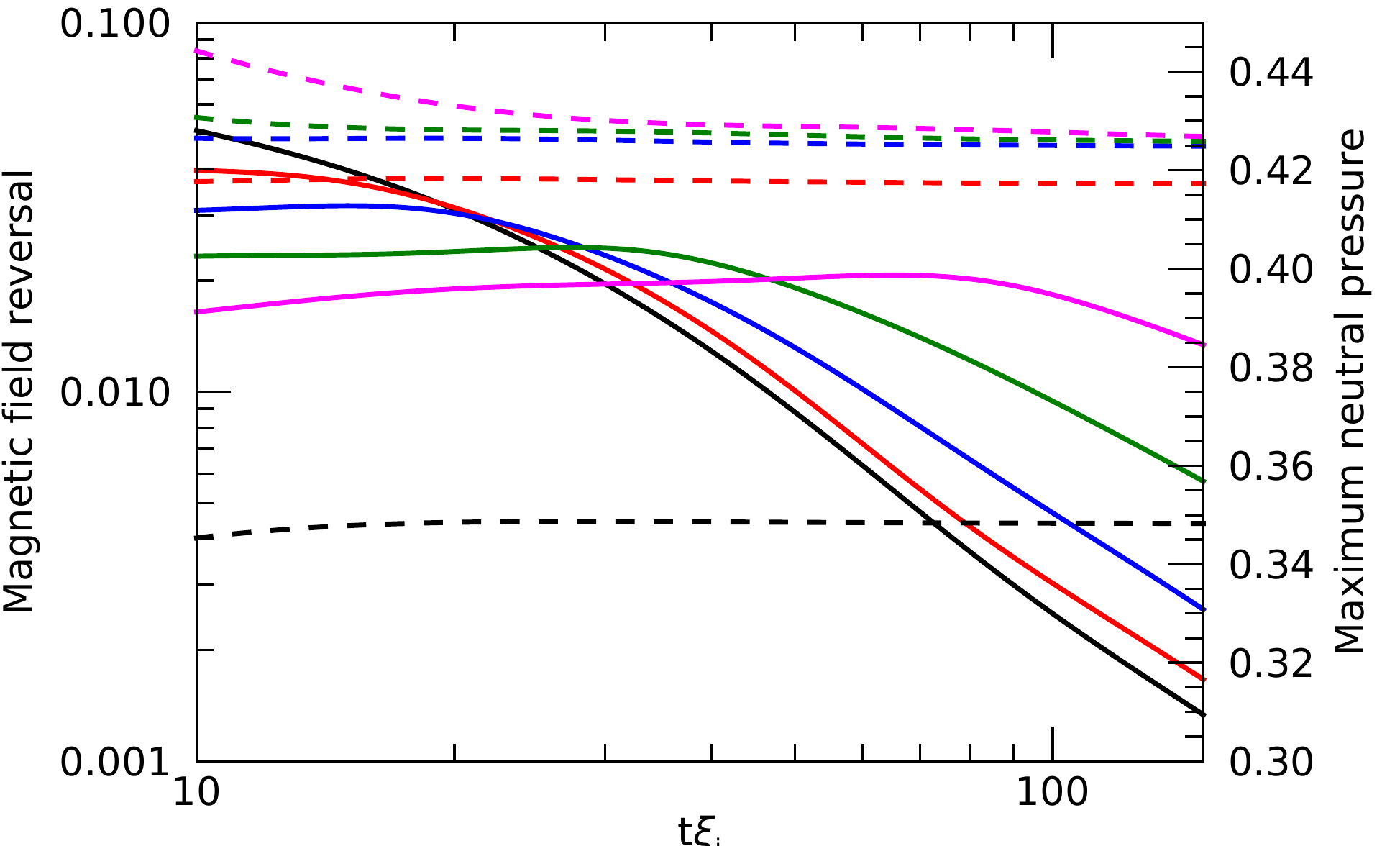}
\caption{(dashed) Maximum neutral pressure and (solid) Magnetic field reversal for different neutral fractions $\xi_{\text{n}}= 0.9 \mbox{ (black),}$ $0.99 \mbox{ (red),}$ $0.999 \mbox{ (blue),}$  $0.9999 \mbox{ (green), }$ $0.99999 \mbox{ (magenta)}$.}
\label{fig:xichange}
\end{figure}

The simulations in this section use $\beta = 0.1, B_x = 0.3 $ with the ionisation fraction $\xi _i$ varying. The results can be scaled by a rough estimate of the time scale changes by multiplying by the ionisation fraction, i.e., $t \xi _i$. As the neutral fraction increases, the influence of the neutral pressure becomes more important. The initial evolution stages generate an overshoot in the neutral pressure due to the collisional coupling. There is then an interplay between the neutral pressure and the magnetic tension. When $\xi_{\text{n}} \leq 0.9$, the neutral pressure increases with time to balance the magnetic tension. When $\xi_{\text{n}} > 0.9$ the neutral pressure overshoot decreases with time. This is shown in Figure \ref{fig:xichange} with the $\xi_{\text{n}} =0.9$ (black dashed) and the $\xi_{\text{n}} = 0.99999$ (magenta dashed) lines. Whilst the two results look very different, the same mechanism is occurring whereby the system seeks an equilibrium; the main difference is the magnitude of the neutral pressure overshoot due to the initial heating. All cases tend towards a constant maximum neutral pressure.   

All neutral fractions investigated feature the intermediate transition as a substructure within the slow mode shock. The normalised time scale does not map the behaviour perfectly and there are differences in magnitude and gradients of the magnetic field across the different neutral fractions. The neutral pressure is the key variable due to its role in the equalising the Lorentz force. As the neutral fraction tends to unity, the maximum neutral pressure is reasonably similar across simulations.




\subsubsection{Changing Beta}

\begin{figure}
\includegraphics[width=0.95\linewidth,clip=true,trim=0.8cm 8.0cm 0.9cm 8.0cm]{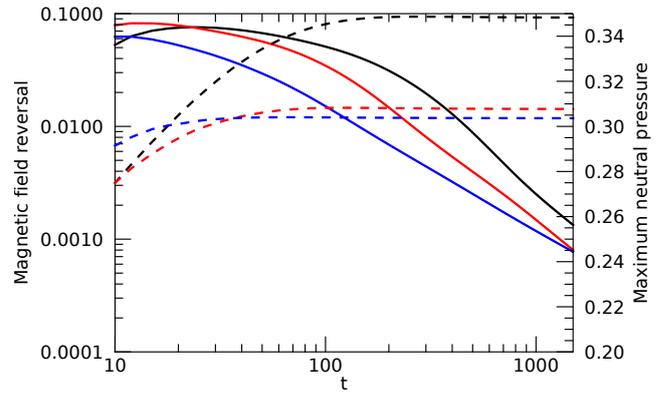}
\caption{(dashed) Neutral pressure and (solid) Magnetic field reversal for different beta values $\beta = 0.1$ (black), $0.01$ (red), and $0.001$ (blue).}
\label{fig:betachange}
\end{figure}


Figure \ref{fig:betachange} shows the neutral pressure and magnetic field reversal for the low-beta regime. All three cases plotted contain roughly the same gradient in magnetic field over time but have different equilibrium neutral pressures. As the plasma-$\beta$ increases, the maximum equilibrium neutral pressure increases drastically. This results in a larger restoring force and the magnetic field decreases more rapidly.

For large plasma-$\beta$ values, a sonic shock can occur in the neutral fluid, as seen in \cite{Hillier2016}. In these simulations, there can be two shocks occurring within the finite-width shock region: intermediate shock in the plasma, and sonic shock in the neutrals.



 
\subsubsection{Changing magnetic angle}

\begin{figure}
\includegraphics[width=0.95\linewidth,clip=true,trim=0.9cm 8.0cm 0.9cm 8.2cm]{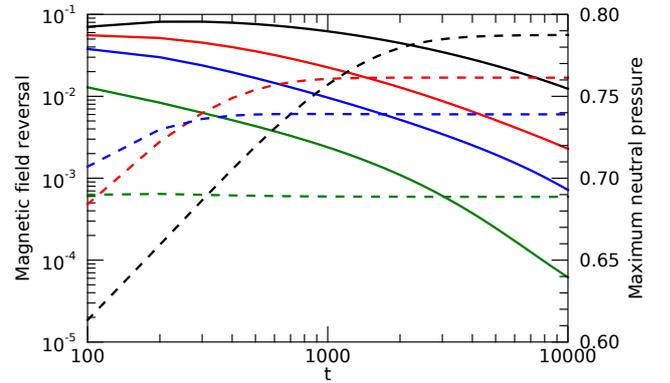}
\caption{(dashed) Neutral pressure and (solid) Magnetic field reversal for $B_x=0.1 \mbox{ (black), } 0.2 \mbox{ (red), } 0.3 \mbox{ (blue), } 0.6 \mbox{ (green)}$.}
\label{fig:bydecay}
\end{figure}

The initial $B_x$ component of magnetic field is modified to alter the angle of the magnetic field. Note that the initial $B_y$ component remains the same and hence the total $B$ strength varies across simulations. Interestingly, the pre-shock region has $|B| \approx 1$ in all simulations as a consequence of the investigated system. 


Changing the $B_x$ value changes the propagation speeds of waves and hence phenomena occurs on different time scales. From Figure \ref{fig:bydecay} it is clear that a larger $B_x$ results in faster equalisation of the neutral pressure and hence the magnetic field reversal and the intermediate shock decay faster, compare to low $B_x$ values. Magnetic tension is the equalising force for the intermediate shock. The magnetic tension increases for larger $B_x$ values and hence the intermediate shock decays faster.





\section{Comments on potential observations}

In the MHD case, the fluid and the magnetic field are frozen together, so in the rest frame of the fluid the electric field becomes $\textbf{E} \propto \textbf{v} \times \textbf{B} = 0$. However, for the two fluid case, there is a separation of the species within the finite-width of the shock.  Within the finite-width, even though there is no electric field felt by the plasma fluid there is therefore an electric field felt by the neutrals. This can be calculated by looking at how fast the neutrals move across the magnetic field, which is the the drift velocity, i.e., $\textbf{E}_{\rm n} \propto (\textbf{v}_{\rm n}-\textbf{v}_{\rm p}) \times \textbf{B}$. This metric is zero either side of the shock since the species are fully coupled. Inside the shock, the ions and neutrals decouple and hence there is a localised increase in the neutral electric field inside the shock substructure. The electric field felt by the neutrals is a potential observable for measuring the neutral-ion drift \citep[e.g.,][]{Anan2014}.

\begin{figure}
\includegraphics[width=0.95\linewidth,clip=true,trim=0.9cm 8.0cm 0.9cm 8.2cm]{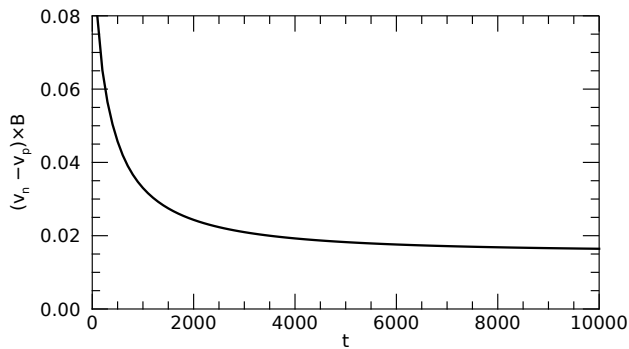}
\caption{Peak neutral electric field increase within the shock through time. Note that the electric field outside the shock in this frame is zero.}
\label{fig:neutralex}
\end{figure}

The maximum electric field felt by the neutrals within the shock is plotted through time in Figure \ref{fig:neutralex} for the reference PIP simulation ($\beta = 1$, $B_x = 0.1$, $\xi = 0.9$). As time advances, this tends towards a constant value indicating that there will always be a localised increase in neutral electric filed within the shock.
The presence of this enhanced neutral electric field could be a potential observable for the effects of partial ionisation in shock waves.

\section{Summary}
This paper has demonstrated that intermediate shocks can occur as substructure inside slow-mode shocks due to partial ionisation. High-resolution 1D numerical simulations were performed to fully resolve the substructure that occurs within the finite width of the slow-mode shock. The physical process involved in forming and dissipating the intermediate shock substructure is as follows: 
\begin{enumerate}
    \item \textbf{Collisional coupling results in overshoots in neutral velocity and density} The initial conditions drive the plasma only. As such, there is a large drift velocity between the two species. The neutral fluids response is to create a Sedov-Taylor-like expansion which creates localised increases in the neutral density and velocity at the interface.
    \item \textbf{Acceleration of the plasma} Our system has velocity driven towards the slow-mode shock at the Alfv\'en speed. Within the finite-width of the shock, the Alfv\'en velocity decreases due to the increased neutral pressure and reversal of the magnetic field, and hence the plasma velocity is accelerated inside the shock to above the Alfv\'en speed.
    \item \textbf{Decay with time} The neutral pressure is not sufficient to balance the Lorentz force and hence there is a gradual evolution whereby the magnetic field reversal tends to zero as time tends towards infinity. The intermediate shock is however sufficiently long-lived to be independent of the initial conditions and considered a physical feature of partially ionised shocks.
\end{enumerate}

The intermediate shock was present as substructure for all tested parameter regimes but the magnitude of the magnetic field reversal was parameter dependent. The larger the equilibrium neutral pressure, the larger the magnetic field reversal. 
As such, the intermediate shock is strongest for high plasma-$\beta$ values.
We would therefore expect this feature to be most significant in the lower atmosphere, e.g., Ellerman bombs. 

The work in this paper is analogous to the magnetospheric intermediate shocks discussed in \cite{Karimabadi1995} where the resistivity creates a dispersive region accelerating the plasma towards the shock region and resulting in an intermediate transition. Here we have a similar effect except it is driven by two-fluid interaction in the absence of resistivity.  

We have shown that an intermediate shock exists for a wide range of parameters, with varying degrees of magnitude and lifetime. Theoretically, this has the potential to be present across a wide range of phenomena in the solar atmosphere where partial-ionisation effects are important, from wave-steepening events (e.g., umbral flashes), to magnetic reconnection (Ellerman bombs, spicules). Future work will be to analyse the implications and observability of intermediate shocks in the lower solar atmosphere.  

In summary, there are four main conclusions in this paper:
\begin{enumerate}
    \item \textbf{Ideal MHD heating across the shock} We have shown analytically that the two-fluid equations reduce to the ideal MHD shock equations when the species are coupled either side of the shock (see Appendix \ref{app:pipht}). The consequence of this is that one would expect to obtain ideal MHD-like heating across the shock, with no additional heating from the collisions.
    \item \textbf{Shocks as substructure in PIP case} Within the finite width of a partially-ionied shock, interactions between the two species can lead to the formation of intermediate transitions within the larger shock structure. Intermediate shocks are a transition from above to below the Alfv\'en speed (here super-Alfv\'en to sub-slow) and feature a reversal in the magnetic field (see Figure \ref{fig:bothref}). This feature forms due to the collisional effects between the two species inside the finite-width shock. An intermediate transition was present for all tested parameter regimes.
    \item \textbf{Potential for large currents} The formation of an intermediate shock features a sharp reversal in the magnetic field across the intermediate shock front, hence there is the potential for large currents to form inside the large-scale slow-mode transition. Large currents are known to play a role in particle acceleration. Hence the formation of intermediate shocks in partially ionised plasma may lead to an additional particle acceleration mechanism.
    \item \textbf{Localised electric field experienced by the neutrals} Within the finite width of the shock, there is a localised increase in neutral electric field. This tends towards a constant value as time tends to infinity, hence may be a potential observable of ion-neutral interactions within shock fronts.
\end{enumerate}

\section*{Acknowledgements}
BS and AH are supported by STFC research grant ST/R000891/1. AH is also supported by STFC Ernest Rutherford Fellowship grant number ST/L00397X/2.


\bibliographystyle{aasjournal} 
\bibliography{losbib} 

\appendix
\section{Hoffman-Teller PIP Equations} \label{app:pipht}

In the Hoffman-Teller frame, the 2-fluid PIP equations are:
\begin{gather}
\rho _{\text{n}} v_{\perp \text{n}} =\mbox{const}, \label{app1:pip1} \\
\rho _{\text{n}} v_{\perp \text{n}} v_{\parallel \text{n}}  = -I_1 +\mbox{const}, \\
\rho _{\text{n}} v_{\perp \text{n}} v_{\perp \text{n}} + P_{\text{n}} = -I_2+\mbox{const}, \\
v_{\perp \text{n}} \left(\frac{\gamma}{\gamma -1} P_{\text{n}} + \frac{1}{2} \rho _{\text{n}} v_{\text{n}} ^2 \right) = -I_3+\mbox{const}, \\
\rho _{\text{p}} v_{\perp \text{p}} = \mbox{const}, \\
\rho _{\text{p}} v_{\perp \text{p}} v_{\parallel \text{p}} - \frac{1}{\mu _0} B_\perp B_\parallel = I_1+\mbox{const}, \\
\rho _{\text{p}} v_{\perp \text{p}} v_{\perp \text{p}} + P_{\text{p}} +\frac{B^2}{2 \mu_0} = I_2+\mbox{const}, \\
v_{\perp \text{p}} \left(\frac{\gamma}{\gamma -1} P_{\text{p}} + \frac{1}{2} \rho _{\text{p}} v_{\text{p}} ^2 \right) = I_3+\mbox{const}, \\
v_{\perp \text{p}} B_\parallel -v_{\parallel \text{p}} B_\perp = 0, \\
B_\perp = \mbox{const}, \\
I_1 = \int \alpha _c (T_{\text{n}},T_{\text{p}}) \rho_{\text{n}} \rho_{\text{p}} (v_{\parallel \text{n}}-v_{\parallel \text{p}}) \mathrm{d} \perp, \\
I_2 = \int \alpha _c (T_{\text{n}},T_{\text{p}}) \rho_{\text{n}} \rho_{\text{p}} (v_{\perp \text{n}}-v_{\perp \text{p}}) \mathrm{d} \perp, \\
I_3 = \int \alpha _c (T_{\text{n}}, T_{\text{p}}) \rho _{\text{n}} \rho _{\text{p}} \nonumber \\ \hspace{0.5cm} \times \left[ \frac{1}{2} (v_\text{n} ^2 - v_\text{p} ^2)+ 3 R_g (T_{\text{n}}-T_{\text{p}}) \right] \mathrm{d} \perp . \label{app1:pip2}
\end{gather}

The integral terms ($I_1,I_2,I_3$) can be removed by adding the neutral and ion equations together, i.e.,

\begin{gather}
\rho _{\text{n}} v_{\perp \text{n}} +\rho _{\text{p}} v_{\perp \text{p}} =\mbox{const}, \label{app1:comb1}\\
\rho _{\text{n}} v_{\perp \text{n}} v_{\parallel \text{n}} +\rho _{\text{p}} v_{\perp \text{p}} v_{\parallel \text{p}} - \frac{1}{\mu _0} B_\perp B_\parallel  = \mbox{const}, \\
\rho _{\text{n}} v_{\perp \text{n}} v_{\perp \text{n}} + P_{\text{n}} +\rho _{\text{p}} v_{\perp \text{p}} v_{\perp \text{p}} + P_{\text{p}} +\frac{B^2}{2 \mu_0}  = \mbox{const}, \\
v_{\perp \text{n}} \left(\frac{\gamma}{\gamma -1} P_{\text{n}} + \frac{1}{2} \rho _{\text{n}} v_{\text{n}} ^2 \right) +v_{\perp \text{p}} \left(\frac{\gamma}{\gamma -1} P_{\text{p}} + \frac{1}{2} \rho _{\text{p}} v_{\text{p}} ^2 \right)  \nonumber \\ \hspace{0.5cm} = \mbox{const}, \\
v_{\perp \text{p}} B_\parallel -v_{\parallel \text{p}} B_\perp = 0, \\
B_\perp = \mbox{const}. \label{app1:comb2}
\end{gather}

The partial pressure and density can be expressed in terms of a total value using the neutral fraction $\xi _n$:
\begin{align}
     \rho _{\text{n}} &= \xi _n \rho _{\text{t}},   \\
     \rho _{\text{p}} &= (1-\xi _n) \rho _{\text{t}}, \\
     P_{\text{n}} &=\frac{\xi _n}{\xi _n + 2 (1-\xi _n)} P_{\text{t}}, \\
     P_{\text{p}} &= \frac{2 (1-\xi _n)}{\xi _n + 2 (1-\xi _n)} P_{\text{t}}.
\end{align}
Substituting these into Equations (\ref{app1:comb1}-\ref{app1:comb2}) gives:

\begin{gather}
\xi _n \rho _{\text{t}} v_{\perp \text{n}} + (1-\xi _n) \rho _t v_{\perp \text{p}} =\mbox{const}, \\
\xi _n \rho _{\text{t}} v_{\perp \text{n}} v_{\parallel \text{n}} +(1-\xi _n) \rho _{\text{t}} v_{\perp \text{p}} v_{\parallel \text{p}} - \frac{1}{\mu _0} B_\perp B_\parallel  = \mbox{const}, \\
\xi _n \rho _{\text{t}} v_{\perp \text{n}} v_{\perp \text{n}} + \frac{\xi _n}{\xi _n + 2 (1-\xi _n)} P_{\text{t}} +(1-\xi _n) \rho _{\text{t}} v_{\perp \text{p}} v_{\perp \text{p}} \nonumber \\ \hspace{0.5cm} + \frac{2 (1-\xi _n)}{\xi _n + 2 (1-\xi _n)} P_{\text{t}} +\frac{B^2}{2 \mu_0}  = \mbox{const}, \\
v_{\perp n} \left(\frac{\gamma}{\gamma -1} \frac{\xi _n}{\xi _n + 2 (1-\xi _n)} P_{\text{t}} + \frac{1}{2} \xi _n \rho _{\text{t}} v_{\text{n}} ^2 \right) \nonumber \\ \hspace{0.5cm} +v_{\perp \text{p}} \left(\frac{\gamma}{\gamma -1} \frac{2 (1-\xi _n)}{\xi _n + 2 (1-\xi _n)} P_{\text{t}} + \frac{1}{2} (1-\xi _n) \rho _{\text{t}} v_{\text{p}} ^2 \right)  = \mbox{const}, \\
v_{\perp \text{p}} B_\parallel -v_{\parallel \text{p}} B_\perp = 0, \\
B_\perp = \mbox{const}.
\end{gather}

Furthermore, we can impose an additional constraint such that either side of the shock the drift velocity equals zero ($v_{ \parallel \text{p} }  = v_{\parallel \text{n}} = v_\parallel$ and $v_{ \perp \text{p} }  = v_{\perp \text{n}} = v_\perp$):

\begin{gather}
\rho _{\text{t}} v_{\perp} =\mbox{const}, \label{app1:fin1}\\
\rho _{\text{t}} v_{\perp} v_{\parallel} - \frac{1}{\mu _0} B_\perp B_\parallel  = \mbox{const}, \\
\rho _{\text{t}} v_{\perp} v_{\perp} + P_{\text{t}} +\frac{B^2}{2 \mu_0}  = \mbox{const}, \\
v_{\perp} \left(\frac{\gamma}{\gamma -1} P_{\text{t}} + \frac{1}{2} \rho _{\text{t}} v^2 \right)  = \mbox{const}, \\
v_{\perp} B_\parallel -v_{\parallel} B_\perp = 0, \\
B_\perp = \mbox{const}. \label{app1:fin2}
\end{gather}

Equations (\ref{app1:fin1}-\ref{app1:fin2}) are identical to the MHD equations except here the density is the \textbf{total} density $\rho _{\text{t}}$ and pressure $P_{\text{t}}$. Therefore, the solution to these equations is identical to the \cite{Hau1989} solution for MHD (Equation \ref{eqn:hau}), independent of the neutral fraction. It should be noted that this is only true over the larger shock structure and inside the shock there is substructure that is highly dependent on the collisional effects. 

   
\end{document}